# Educational Architecture and Emerging Technologies: Innovative Classroom-Models.

*Arquitectura educative y Tecnologías emergentes: Modelos de aula innovadora.*

José Gómez Galán. *University of Extremadura (Spain) & Metropolitan University-AGMUS (Puerto Rico –USA).*

Contacto: jgomez@unex.es / jogomez@suagm.edu

**ABSTRACT**
This article has as its main objective to emphasize the importance that physical spaces and architecture should have for a correct pedagogical integration of the ICT, including in this concept all the current means of communication and, in particular, the emerging technologies. Not only is it because, as has been demonstrated, learning environments have a decisive impact on teaching processes and students' own learning, but it is also imperative that they be adapted for digital literacy, inevitable today in a society dominated by ICT. The use of post-structural theoretical methodology offers an analysis of the current situation in relation to immediate educational needs, obtaining as a result that a redefinition of physical learning spaces is urgently needed if we really want to carry out an education that responds to the social impact of emerging technologies. In this sense, and focusing on the physical classroom or traditional classroom as the basic space unit of the centers - and in general of the educational system on the whole - which has practically not changed its traditional structure, one points towards an innovative model of the more versatile and flexible, adapted to the current needs of a new curriculum and a new education for the citizens of the 21st century.

**KEYWORDS**
Educational Architecture, Emerging Technologies, Physical Learning Spaces, Digital Society, Physical Classroom.

**RESUMEN**
Este artículo tiene como objetivo principal destacar la importancia que en la actualidad deben tener los espacios físicos y la arquitectura educativa para una correcta integración pedagógica de las TIC, incluyendo en este concepto a todos los medios de comunicación actuales y, en especial, a las tecnologías emergentes. No sólo es debido a que, como ha sido demostrado, los entornos de aprendizaje tengan un decisivo impacto en los procesos de enseñanza y en el propio aprendizaje de los estudiantes, sino que resulta obligatorio que estén adaptados para una alfabetización digital hoy inevitable en una sociedad dominada por las TIC. Mediante el empleo de una metodología teórica post-estructuralista se ofrece un análisis de la situación actual frente a las necesidades educativas inmediatas, obteniendo como resultado que urge una redefinición de los espacios físicos educativos si realmente queremos llevar a cabo una educación que responda al impacto social que implican las tecnologías emergentes. En este sentido, y centrándonos en el aula física como unidad espacial básica de los centros -y en general del sistema educativo- que prácticamente no ha visto modificada su estructura tradicional, se propone finalmente un modelo innovador de la misma, más versátil y flexible, adaptado a las necesidades actuales de un nuevo currículum y una nueva educación para los ciudadanos del siglo XXI.

**PALABRAS CLAVE**
Arquitectura Educativa, Tecnologías Emergentes, Espacios Físicos de Aprendizaje, Sociedad Digital, Aula Tradicional.

## 1. INTRODUCTION

One of the aspects which best reflects in what way teaching-learning processes are developed in the classroom has as much to do with the way space is distributed as the type of material used or the academic furniture, and this could apply to centers of education in general (from infant education up to and including education in universities) and to the particular philosophy employed or educational project that sustains the didactic action which develops in relation with distinct installations, the way in which they are distributed and the amount and quality of equipment which they have at their disposal. The progressive development of the use of information and communication technologies (ICT) will permit increasingly richer and more efficient contexts from a pedagogical viewpoint, making learning by discovery possible (López Meneses & Gómez Galán, 2010; Gómez Galán, 2011 and 2014).

It is evident that whatever person, including those who are unconnected with the educational world in a professional way, with only a superficial glance at an image of a classroom which belonged to another epoch would notice immediately, because of the position and type of school desks, the disposition and characteristics of the space available and the type of scholastic material, the type of methodology the teacher would have employed and indirectly would be able to have an idea of the type of subject-content to be taught- in many cases only transmitted with these methodologies and materials which are out of date, and of course the type of teacher-pupil and pupil-pupil interaction would be explicitly clear.

This concerns, and it is necessary to keep this in mind, curricular elements which are decisive in all teaching-learning processes regardless of whether this means primary, secondary or higher levels of education, and we can affirm that there is a peculiar interdependence between all of these links or elements mentioned above and the availability and distribution of space, materials, furniture and resources, up to the point in which practical learning can see itself conditioned by these elements, which shows their relevance in school dynamics (Gómez Galán & Mateos, 2002).

It is due to this that given the responsibility that they have concerning the other elements of the curriculum (objectives, content, pedagogical methods, evaluation criteria, age groups, cycles, grades, teaching practice, etc.). The distribution of space within the classroom, the arrangement of furniture and the materials-resources has to be appropriate for the obtaining of educational objectives which are set, and these objectives are obviously shaped by the type of education required by the new social realities dominant in each moment. And as is natural, the demands of our technological society, in the 21st century, are starting to shape new educational models which require new, well distributed space and endowments.

In the present day society in which we live, commonly called the information society, because of the irruption and part played by in all areas of our lives by ICT (Castells, 2000; Beniger, 2009; Webster, 2014; Hofkirchner & Burgin, 2017). In the education institution there are multiple and continuous initiatives undertaken in order to introduce and integrate these new technologies into schools, so as to explore and pursue all the possibilities of their application to education in the distinct educational levels. The new objectives, content, materials, etc, which are needed for an adequate formation in technology and means of communication, require new planning of space in classrooms, an essential aspect which cannot be ignored (Mateos & Gómez Galán, 2002; Gómez Galán, 2015a).

## 2. DESIGN A NEW MODEL OF CLASSROOM: THE FUTURE OF EDUCATIONAL ARCHITECTURE

It is imperative that a new type of physical environment is designed for use in formal education (Skill & Young, 2002; Roberts & Weaver, 2006; Brooks, 2011; Carroll, 2015). The evidence suggests that enhanced learning environments, independent of all other factors, have a significant and positive impact on student learning ((Zendan, 2010; Blackmore, Bateman, Loughlin, O'Mara & Aranda, 2011; Brooks, 2011; Ariani, 2015; Byers, 2016; Barrett, Davies, Zhang & Barrett, 2017). Sometimes, it would be necessary for centers to be newly constructed, when there is no other way of supplying for serious deficiencies, but in most cases this can be achieved through adapting the spaces and location of equipment; only in this way will it be possible to offer quality education in accordance with the needs of students and the demand or requirements of our technological society.

Of course, if they endeavor to reach these necessary goals not only would they require new types of environment. This would be one of the pillars that would sustain the quality of today's education, but not the only one. Above all we should begin with the need to form- initially and continually-teachers who face this educational challenge from a technical and above all pedagogical angle (Gómez Galán, 2002 and 2015b; Miranda & Gómez Galán, 2016). It would not be of much use if, after disposing of conditioned spaces for the employment of ICT and audio-visual media, the teachers did not have even a basic knowledge of the didactic use of the main technological instruments and or their correct application in the teaching-learning process.

Nevertheless the element that we are considering, the use of educational space remains fundamentally important. Well, naturally, because even if one obtains teaching personnel who are sufficiently qualified to educate in and with available technologies, at this present time school spaces are not yet adapted to fit in with the technological needs of today. It is therefore necessary, more than ever, that architectural designs used for building schools becomes a labor which is carried out by professionals working in collaboration, who place pedagogical requirements above and beyond aesthetic or traditional considerations.

In this new context, education has to offer responses to new needs. The human being of the 21st century needs knowledge, capabilities, and outlooks for a new society in which the communicative processes have so much relevance. It is for this reason that we are currently speaking insistently of the need for digital literacy. Just like traditionally, when one of the basic objectives of education was linguistic and cognitive literacy – with the acquisition of fundamental competencies such as, for example, the ability to read and to write –today, literacy in the new multimedia and hypermedia languages, born in the digital revolution, are also necessary. And the necessary capabilities for using the new tools that make their use and development possible. What we would definitively call digital literacy.

Currently there is abundant scientific literature which attempts to propose answers to what could be the main directives for carrying out adequate digital literacy. There are some important recent contributions according to the educational level of interest, such as Casey & Hallissy (2014), Ainley, Schulz & Fraillon (2016), Reynolds (2016), or Scarcelli & Riva (2016). Also to be mentioned is the works of Littlejohn, Beetham & McGill (2012), Gillen (2014) and Hagerman & Spires (2017) for establishing the state of the question and new approaches to digital literacy.

We consider, however, that on too many occasions the trees don't stop seeing the forest, and in too many research figures from the international scientific community the problem itself certainly appears limited, if only from an instrumental and technical perspective. Digital literacy is reduced to try to develop didactic strategies for the acquisition of competencies in the management of computers, digital whiteboards, tablets, smartphones, operating systems, information searches on the Internet, etc., that is, basic training in hardware and software from a primarily operational dimension.

But the problem today is much greater and should be confronted from its roots. The majority of the school-age population con-tinually acquires digital capabilities in their daily lives, since ICT dominates everything. Except in the cases of digital literacy in the adult population, the didactic instrumental focus is not necessary. The children and young adults of the communication society are perfectly familiar with the employment of new media. The problem, without a doubt, is something else: that these new media decisively condition their lives, they influence them, they affect them, they guide their habits, their interests, their opinions, etc. They are definitely omnipresent in their world; they are their world. And what results is important to highlight: in some way, they are the "new media." As such, the educational strategies that we need should be consistent with this and face the real problem.

We will explain in detail the reason for all of this. We start from the fact that we have developed widely in other contributions (Gómez Galán, 2003, 2011, 2014 and 2017), that at present we are, due to the digital revolution created by information and communication technology (ICT), in a process of convergence that we call techno-media, and in which media –as much the traditional as the newest, from the press or the radio to the Internet and social networks- stop existing as separate entities in order to form part of a unique digital media that covers the whole of human communication and in which, without a doubt, we find ourselves at its dawn. In the future, there will be only one medium –with many manifestations, naturally, but participating as one essence– in which we will not be able to speak of distinct communicative and/or technological confines. Watching television will mean interacting with what today we do with social networks, reading the press at the same time that we create our own content, listening to the radio while we envision the participation of talk show guests, and we will communicate with each other through videoconferencing with them, etc., the divisions between media will not exist and we will find ourselves in a unique interactive communication system, born of the digital paradigm.

It will even create a convergence of hardware and software, and we will not have a variety of them available to us for the transmission of the different media-technologies, the traditional ones, and we'll call them the new ones. The medium-system will be the same, the only thing that will change will be the size of this medium that we will have at our disposition, through the Internet/Digital paradigm, and we will have all the existing media. The only difference will be the comfort of viewing, portability, etc. But they will have the same functions. That is, we will not talk about different media, but rather of one medium with different sizes adapted to our needs.

But in this new reality, let's not trick ourselves, the structure of power –economic, social, political, etc. – of the classic communication media will not disappear. They will just have been adapted to a new technological framework, but they will continue exercising their influence and domination over citizens, even more intensely than currently, since they will be offering an apparent -yes, apparent- freedom to the user-consumer-voter, who, thanks to the development of digital technologies, will believe that they have widely gained independence and decision-making. But this is just an illusion created by means of super-sophisticated communication channels that are controlled, without a doubt, by the same powers that be as always. For this reason, more than for a society of information or communication – in that the term might denote complete freedom in the flows of communication– it would be necessary to talk of a techno-media society. A society in which the digital revolution, in some way as a result of an adaptation, in the continual processes that the history of humanity supposes, of the power structures to new scenery in which physical space stops being decisive. And which, as always, supposes enormous benefits in all fields: economic, social, power.

The educational problem of forming our future citizens continues, therefore, including the new situation that emerged in the 20th century at the start of the emergence of mass communication media. The digital revolution has promoted its effects enormously, since it has brought the power of communication and its impact on society to extraordinary limits. But this is not a completely new problem. It's the same problem, but increased. And what needs to change is the authentic challenge of digital literacy. In short, we need new educational spaces for the new digital literacy. But should not be spaces to integrate only computers, digital whiteboards, tablets, smartphones, etc. This will be a mistake. We need new spaces to integrate media education in this global context. The educational institutions are in need of infusion of new model of classroom in order to integrate a new curriculum. A new architecture is needed to discover a new education.

### 3. PHYSICAL LEARNING SPACES FOR EMERGING TECHNOLOGIES: A GLOBAL PROPOSAL

New spaces which are more versatile and flexible than those in use at the moment are required, spaces which offer the possibility of integrating adequately emerging technologies and the media, which can be employed with efficiency within the architectural environment. And although this is not a new problem, and some studies written many years back ago argued for the introduction of computers into educational establishments from the perspective of the learning environment (Hiemstra, 1991; Nelson & Sundt, 1993; Martin-Moreno, 1994; Zandvliet & Straker 2001; Zandvliet & Fraser, 2004), nowadays it is difficult to condition school spaces in response to the inrush of technology, fundamental in the formative process of students in different educational stages.

If in fact we can consider the classroom as the basic spatial unit of the centre, and we will focus on it, it would be opportune to say that all the ideal characteristics that we suggest (and that occasionally it will not be possible to achieve them, fundamentally due to economic reasons) can be applied to whatever educational space (classrooms, halls, workshops, laboratories, seminar rooms, etc. For the use of new media and emerging technologies - based on traditional, in the techno-media convergence context is one only (Gómez Galán, 2015) - it is necessary to fulfill the following conditions:

- Spaces must be ample, but at the same time be able to offer top quality images and an ideal sound transmission throughout the auditorium.
- Furniture can be moved easily, with the aim of applying new didactic methods - such as group dynamics- which are essential when working with ICT.
- Students and teachers will be comfortable, stress free, with space being one more pedagogic resource, serving education not hindering it.
- There will be special spaces for some activities, although from the point of view of the application of information and media technology it is to be recommended that these can be developed in whatever classroom, which can be conditioned for it.
- It is necessary that there be a complete integration between traditional elements and those contained within emerging technologies; so for example, we are not speaking about converting the classroom in a fully modernized classroom, although nowadays this is very important, but about the need of teachers to have available emerging terminals that enable students to access information through the internet, in distinct databases and multimedia and virtual contexts.
- New spaces must avoid, in the best way possible, the nuisance of external noises, or noise from nearby classrooms (it will produce more noise than that generated by traditional media, and it is necessary to counteract it by isolating acoustically the study areas.
- There will be different workshops and laboratories for those subjects (fundamentally the optional ones) which require special instruments, this is fundamental in concrete educational stages, such as secondary education or university education.
- We should offer, above all, quality spaces, that not only avoid adding difficulties to didactic and educational processes, but that moreover, they help continually towards improving them. Within this philosophy, for the equipment and material it is preferable to opt for, when possible, the best quality. In some educational stages, whether secondary or university education, the durability of the equipment is fundamentally important, and it is something we should bear in mind (because all material used in schools wears out quicker). In the case of nursery or primary education durable equipment would be an essential requirement.

All this will be accompanied by different strategies that help to improve the pedagogic framework within which scholastic work is developed. We are in complete agreement with those put forward by Martin-Moreno (1994) that in our case, we would fundamentally apply for the use of new media in the centers. These are:

- A search for a correspondence between building and furniture designs.
- Development of a new conception of the student work-place.
- Redistribution of work areas through mobility of the furniture.
- Student participation.
- Consideration taken of urban dimension.

The architectural changes and spatial innovations that we consider fundamental for the integration of ICT in educational establishments, from traditional architecture, should be as follows (Gómez Galán, 2001, 2002 and 2014; Mateos & Gómez Galán, 2002):

- The larger size of the classrooms stands out most of all, and they are moreover conditioned for the use of audio-visual media (they isolate sound, with light sources placed strategically, etc.)
- The creation of a technological classroom, that can be also used like the old auditorium but prepared for the ubication of most of the technological instruments, and in which they can be used, or the modification of the library as a resource centre and data base. These changes, moreover, offer us a new architectural form of the building.
- The technological classroom will be equipped sufficiently so as to be able to offer all the instruments necessary for the use of the emeging technologies and the audio-visual media in the centre. It is evident that it will be here where all the technological material will be housed, and the technology can be used there or moved to the classrooms or halls where its use is pertinent (of course there will be instruments which will be permanently kept in the technological classroom, whether due to its excessive weight, fragility or high price).

Without claiming to be exhaustive (and avoiding all reference to specific to capacity and velocity, so as to avoid dating this chapter) we think that a centre of primary or secondary education should have the following technological possession (with sufficient workspace in the technological classroom for 30 students), keeping in mind the fact that this will depend on economic possibilities, the needs or formation of teaching staff (it would be absurd to have available, for example, really digital cameras expensive or 3D virtual reality systems if the teacher does not know how to carry out a basic audio-visual set-up; We insist once again in the need of formation of teaching staff in the new technologies):

**Hardware**
- Information technology and telematics: 15 multimedia personal computers (with the most recent processor; generous RAM and graphic card memory, large hard disk, if possible SSD; large monitor with high resolution; multimedia system;

keyboard and mouse with economical design and high performance,; the format of the box should be tabletop or semi-tower). 15 portable computers and 30 tablets (if funds allow, of the highest performance possible). 7 printers (photographic quality and high volume of prints; bearing in mind the running costs and choosing the most economical version) with tabletop scanners (high capacity); 30 e-book readers, connection of all systems to internet; new 3D virtual and augmented reality systems (3D video projection systems, 3D displays, stereoscopic screens, virtual rooms, autostereoscopic displays, holograms, simulators, haptic and feedback equipment - VR glasses, chairs, gloves, etc,, which involves student's immersion of "virtual room", "virtual environments" and "tele-immersion" type-, etc.); 3D scanning and printing devices.

- Audio-visual and communication media: 1 portable video projector of high resolution; 1 portable retro-projector; 1 digital board; 1 electrical projection screen; 1 edition table; 3 television sets (as big as possible, with program synthesiser and full remote control); 2 video cameras (with tripod); 1 Blue-Ray package with virtual digital surround; 1 presenter of documents in 3D (connectable to monitor, videoprojector and printer; copying of solid and photographic items); 1 full set of sound equipment, with possibilities of radiophonic production.
- Furniture and general infrastructure: tables for computer equipment; meeting tables; metal security cupboards for technological instruments; desks; portable projection screens, shelves, shafts for ventilation and safety cables; traditional green blackboard for chalk and whiteboard for felt pens; chairs (on wheels as the tables); etc. Biotechnology must dominate in the furniture, especially concerning the seats.
- In all the technological instruments presented (we insist again that it is just a possible list to give you an example of what may be needed) care will be taken to ensure that the least possible number of decibels of sound be emitted. This should always be a very important factor to consider. If given the choice, always choose the most silent apparatus.

It must also be said that this list is appropriate for developed countries. In countries with greater economic difficulty the requirements of course would be greatly reduced. However, it is important to state that in order not to miss out on the wake of the revolution of communication and information technologies, it is essential to confront the teaching-learning of computers, telematics and audio-visual media from the school. Nevertheless, it is not necessary to rely on the most up to date technological instruments to establish the foundations of an adequate and effective training for today's needs. What is important is to offer these instructive procedures.

**Software**
- Of course, it will be necessary to have all the appropriate software for a full educational performance of the equipment previously mentioned. Naturally, and in the case of computers or tablets, we consider the presence of office computer systems packages (word processors, spread sheets, data bases, presentations), anti-virus, internet utilities (browsers, E-mail functions, news blocks, web page editors, firewalls etc), auto editing programs, graphic design, photographic imaging, sound and video, system utilities, etc. Moreover of the specific educational software: general and thematic encyclopaedias, programs about curriculum contents, library functions, data functions, scholastic functions, etc. Nevertheless, the fundamental decision falls once more upon the operative system (that moreover can affect decisively the total price of software). In particular we recommend the use of Android and Linux (depending on which device) for a number of reasons, although in the case of centers in which teaching staff are accustomed to the use of commercial operative systems, one could propose this option in accordance with estimated possibilities.

- This is precisely, today, the main reason to opt for GNU/Linux, besides the fact that it is a quality operative system, which is trustworthy and reliable: its free distribution together with its applications means no software costs. Originally, the main disadvantage in using Linux was its relatively major complexity when used in relation to some other commercial operative systems; nevertheless, today this inconvenience does not exist. Studies made about this really demonstrate that the saving made by not having to pay licensing costs can be spent on adapting infrastructures (that in the case of education would suppose versatile spaces for the use of emeging technologies) and the stocking up of hardware. It is estimated that the elimination of proprietary software costs would double the amount of computers bought (De Icaza, 2002), which would reduce the student-computer ratio by half. This would reduce other costs, eliminating the use of proprietary client software and reducing maintenance and additional support expenses (Díaz, Harari & Banchoff, 2005; Stallman, 2009; Duque, Uribe & Taberes, 2016). All in all this is could be an interesting alternative for developing countries. Leiva & Moreno (2015, p.1) argue that "it is necessary to set up a new model of school committed an educational, intercultural and inclusive perspective on the implementation of proposals and initiatives based on the equipment of workplaces with low-cost equipment and the design, installation and development of open source software".
- With regard to the software, finally, it is worth pointing out that the educational stages would condition the needs that must be supplied, so for this we present the guidelines we have laid out as global thus avoiding an excessive specification.

## 4. CONCLUSIONS.

Reflecting what is happening in society, in a historical period characterized by information and communication processes, experiences centered on taking advantage of what new technology can offer us is acquiring a leading role in educational science. The Education cannot constitute a reality apart or separate from the other social systems. Besides making use of the different technological instruments to optimise, as far as possible the teaching learning processes, the educators need to form their students to know more fully the world in which they live, to reflect critically on their own existence and to contribute towards improvement and progress. And in this context, the educational physical spaces are essential.

Numerous specialists in educational technology argue for the integration of ICT in the classroom teaching-learning process, at different educational levels (Bernal & Rodríguez, 2009; Léger, 2014; Cobos, Gómez Galán & López Meneses, 2016; Martín Padilla & López Meneses, 2016; Roig-Vila, 2016). Nevertheless, there are very few studies which have stressed the importance of adequate layouts and educational space for classrooms that bring together the best conditions for the correct use of hardware (Bautista & Borges, 2013; Yang, Huang & Li, 2013; Ellis & Goodyear, 2016; Gormley, Glynn, Brown & Doyle, 2016). It is necessary to propose a new type of educational architecture, which is more adequate than the traditional, so as to be able, as much as possible, and to use ICT in the most effective, pedagogical way (Gómez Galán & Mateos, 2002).

In the face of this new digital reality, we believe that an evolution of the methods and didactic models is necessary, starting from the traditional bases, which support it. For this purpose we need new spaces with full integration of ICT. One of the strategies that we consider most adequate, that would be only an example, is precisely the dialogue within educational technology, about combining within the processes of teaching-learning the presence of the media with didactic resources—all citizens, regardless of age, are users and consumers of them—as an object of study. The facilities that the digital paradigm offers us for the use of the media in the classroom grant us a decisive advantage in order to strengthen education in it.

Let's look at an example. Let's think, without going any further, about the possibilities of the press. It would not be the same to have a newspaper available in the classroom with which to develop a concrete didactic activity, as accessing editions on the Internet, which are permanently updated, of four or five newspapers, to take advantage of all the possibilities that they give us, in order to, let's state our case, do a critical analysis of the news (reading the same news in different papers, measuring the space dedicated by each of them, the treatment, utilized sources, and contrast among them, etc.) The Internet, it's clear, would have made this task easier (and in the opposite way we would have needed to acquire multiple issues, to highlight, to trim the content, etc.) but, in essence, the process of study and pedagogical use of ICT –today turned into powerful media—will be able to support itself on the classical structure of media education (Gómez Galán, 2011 and 2015).

We can make all this extend to any computer tool or new telematic tool, for any emerging technology that we think, not only the present ones but also the future ones. We could utilize any of them for didactic development with the goal of strengthening interactivity, group work, creativity, motivation, etc., and surely the results—if everything has been adequately designed based on pedagogy—will be positive. However, and as we widely defended on another occasion (Gómez Galán, 2011), we cannot stop the margin of impact that these have on society, and especially their influence on childhood and youth. We should take advantage of their use as didactic resources in order to turn them into parallel objects of study, of analysis of their characteristics, of the use that the student makes of them in his life, of the advantages, but also possible dangers that they contain. It would even be adequate to first employ technologies under educational control as preparation for the knowledge and management of the most socially extended, of which, without a doubt, children and young adults will turn into, sooner or later, and then into consumers. We would talk about, therefore, an integral education in which ICT/media is contemplated in their authentic social and educational dimension, in the sphere of techno-media convergence.

In the present technological society to educate students in information and communication technology, within the education system, managing to integrate these tools and making them an object of study in the classroom, demands of teaching professionals to reconsider their conception of scholastic space and condition them for the optimum development of the educational function, so that the methodological and organizational aspects arrive to play a necessary and decisive role, which has not been done sufficiently. The pedagogic use of technological instruments involves a profound need of a determined infrastructure, and the modification of the position of information and communication systems and audio-visual and multimedia equipment in accordance with educational criteria. Of course not only is an adaptation of space needed for the new emerging technologies, but also an optimum furnishing -and adapted to formative needs- of hardware and software, as well as sufficiently qualified teachers for this task.

.

## 4. REFERENCIAS BIBLIOGRÁFICAS